\documentclass[%
reprint,
superscriptaddress,
showpacs,preprintnumbers,
amsmath,amssymb,
aps,
pra,
]{revtex4-1}
\usepackage{dcolumn}
\usepackage{float}
\usepackage{graphicx}
\usepackage{dcolumn}
\usepackage{bm}
\usepackage{epstopdf}
\usepackage{mathtools}
\usepackage{physics}
\usepackage{xcolor}
\usepackage{hyperref}

\begin{document}

    \title{Optical responses of Fano resonators in non-spectral parametric domains}
    %
    
    \author{Ankit Kumar Singh}\email{AnkitKumar.Singh@leibniz-ipht.de}
    \affiliation{Leibniz Institute of Photonic Technology, Albert-Einstein-Straße 9, 07745 Jena, Germany}

    \author{Jer-Shing Huang}\email{Jer-Shing.Huang@leibniz-ipht.de}
    \affiliation{Leibniz Institute of Photonic Technology, Albert-Einstein-Straße 9, 07745 Jena, Germany}
    \affiliation{Institute of Physical Chemistry and Abbe Center of Photonics, Friedrich-Schiller-Universität Jena, Helmholtzweg 4, D-07743 Jena, Germany}
    \affiliation{Research Center for Applied Sciences, Academia Sinica, 128 Sec. 2, Academia Road, Nankang District, Taipei 11529, Taiwan}
    \affiliation{Department of Electrophysics, National Yang Ming Chiao Tung University, Hsinchu 30010, Taiwan}   
    %
    
    
    \begin{abstract}
Fano resonance observed in various classical and quantum systems features an asymmetric spectral line shape. For designing nanoresonators for monochromatic applications, it is beneficial to describe Fano resonance in non-spectral parametric domains of critical structural parameters. We develop the analytical model of the parametric Fano profile based on a coupled harmonic oscillator (CHO) model and theoretically demonstrate its application in describing the optical response of a chirped waveguided plasmonic crystal (CWPC). The developed parametric Fano model may find applications in the design of  monochromatic and spectrometer-free nanodevices.

    \end{abstract}
    
    \maketitle
    
    
    \section{Introduction}
    
The distinctive asymmetric spectral line shape observed due to spectral interference of a discrete and a continuum mode in the coupled resonator system is known as the Fano resonance. It is widely observed in various quantum and classical optical systems \cite{fano1961effects,luk2010fano,limonov2017fano,miroshnichenko2010fano,wu2012fano,chen2021spectrometer,zhang2009manipul-bowtie,ott2014prl}. 
While most of the works deal with the asymmetric spectral profiles of a single resonator in the frequency domain, Fano profiles can also be observed if the optical response of a single frequency from a Fano resonator is mapped under gradually tuned excitation polarization \cite{ray2017polarization,valentim2013asymmetry}, varying system temperature \cite{zhang2018thermally,zheng2017compact} or in a series of Fano resonators with gradually varying structural parameters at a single frequency \cite{chen2021spectrometer,see2017design,ouyang2020spatially,ray2019controlling}. In this case, Fano resonance manifests as asymmetric intensity profiles with respect to experimentally accessible parameters. Such Fano-like asymmetric profiles in non-spectral parametric domains, which we called ``parametric Fano'' profiles, are useful for designing Fano resonators for monochromatic applications because they provide direct information about the dependence of optical response at the operational frequency on the critical structural parameters. For example, in nanooptics research, it is very common to design and fabricate a series of similar nanostructures with gradually varying dimensions in order to obtain the best nanodevice that gives the highest signal of interest under a monochromatic excitation \cite{ouyang2020spatially,tittl2018imaging,chen2014modulation}. In addition to the structural parameter space, the tuning of optical response is also seen in the temporal domain for non-linear optical systems \cite{fardad2019parametric}. Therefore, the ability to describe Fano profiles in non-spectral parametric domains allows researchers to understand the impact of specific structural parameters on the optical response and significantly increases the chance of obtaining nanostructures with the desired optical properties. However, to the best of our knowledge, an analytical model to describe such a monochromatic optical response of a Fano resonant nanostructure is still lacking.

In this work, we develop an analytical model to describe the linear optical response of a series of plasmonic nanoresonators as a function of a specific structural parameter. The approach reduces the effort of scanning structures in order to obtain the best structure for a desired optical property. The model is based on the coupled harmonic oscillator (CHO), where the monochromatic optical resonance is evaluated based on the structural parameters of the system rather than the frequency. We obtained parametric Fano resonance formula to model the monochromatic optical response of a chirped waveguided plasmonic crystal (CWPC) in parametric domain and estimate the parameters of the two participating modes. The resonance profile observed in
the parametric domain shows similar asymmetry, steep edges, and dispersive behavior as in the spectral domain. The presented approach allows describing monochromatic optical resonances of nanoresonators, including Fano resonances, in non-spectral parametric domains and may facilitate the design of nanostructures for monochromatic applications, for example, spectrometer-free sensing and nonlinear signal generation using an excitation at a fixed frequency \cite{chen2014modulation}.
    
\section{Illustration of parametric resonance on geometrical parameter space}

\begin{figure*}[ht]
	\centering
	\includegraphics[width=0.8\linewidth]{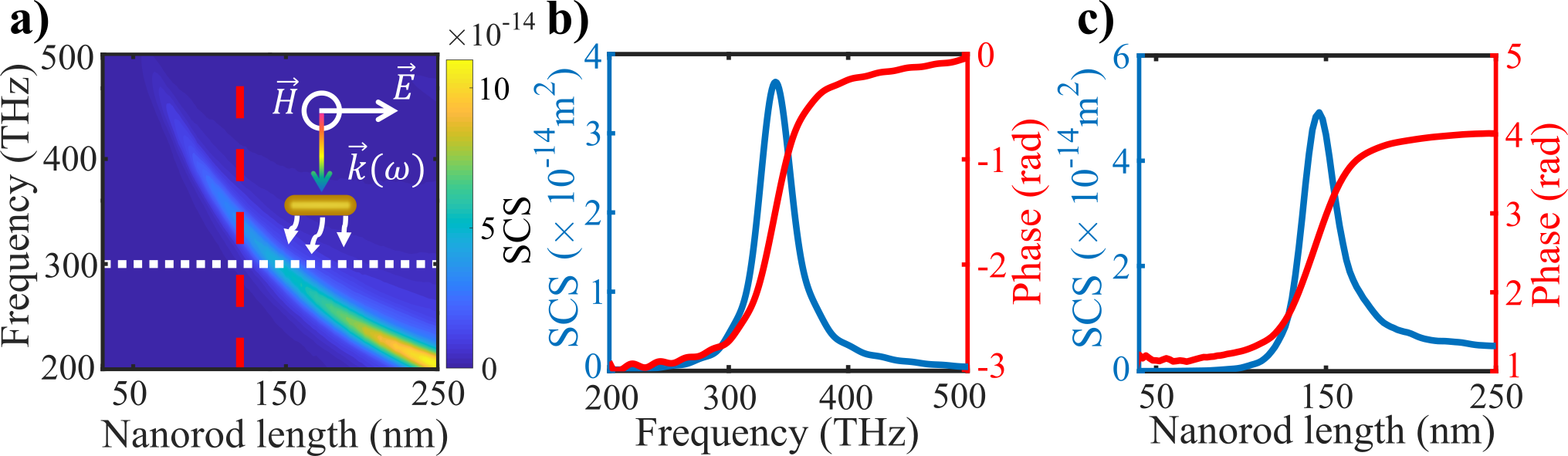}
	\caption{\label{fig1}Spectral and parametric resonances of a series of gold nanorods in water. (a) The simulated SCS (color bar) of gold nanorods on a two-dimensional plane of the excitation frequency (vertical axis) and the nanorod length (horizontal axis). The inset shows an exemplary case of scattering geometry of the nanorod of fixed length. (b) Frequency-dependent SCS (blue, left axis) and phase (red, right axis) of a nanorod with a fixed length of 120 nm (along the red dashed line in a). (c)  Length-dependent SCS (blue, left axis) and phase (red, right axis) of a series of nanorods at the excitation frequency of 300 THz (along the white dotted line in a).}

\end{figure*}

In the following, we illustrate the concept of ``parametric resonance'' dependent on structural parameter by showing the resonant amplitude of nanorods as a function of the rod length. We theoretically study the longitudinal scattering cross-section (SCS, incident light polarized along the long rod axis) of a gold nanorod in water with a fixed cross-sectional radius of 15 nm and different nanorod lengths. SCS is obtained from numerical simulation based on finite-difference time-domain method (FDTD Solutions, Lumerical, see Appendix) and the dielectric constant of gold is obtained from experimental data \cite{johnson1972optical}. The nanorod is meshed with a cubicle meshes of size 1 nm$^3$. Fig.\ref{fig1}a shows the SCS of different nanorod lengths mapped to a two-dimensional space of excitation frequency (vertical axis) and nanorod lengths (horizontal axis). While the vertical line-cut profile (red dashed line) shows the typical scattering spectrum of a nanorod of a fixed length, the horizontal line-cut profile (white dotted line) shows the ``monochromatic'' SCS as a function of the rod length. The latter shows the ``parametric resonance'' of the nanorod, as it is the dependence of monochromatic optical response on a structural parameter (the nanorod length). The spectral resonance of a nanorod with a length of 120 nm (obtained from SCS along the red dashed line in Fig.\ref{fig1}a) is shown in Fig.\ref{fig1}b, where SCS and phase are plotted as a function of the excitation frequency. This is a typical frequency-domain spectrum of a plasmonic nanorod with clear features of a Lorentzian line shape and, a $\pi$ phase shift over the resonance in the electric field evaluated at the center of the particle. If we now look at the monochromatic optical response of a series of nanorods, Lorentzian-like profiles of SCS and phase are observed as the nanorod length is increased. Fig. \ref{fig1}c shows SCS and phase of different nanorods with increasing length, excited with a source fixed at 300 THz (obtained from SCS along the white dotted line in Fig. \ref{fig1} a). A resonant enhancement is obtained at the length of 146 nm. The phase obtained from the electric field shows a phase shift of about $\pi$ relative to the excitation field over the resonant rod length. The phase gain across the resonance is associated with the Lorentzian-like resonance in the SCS, whose  length  is inversely proportional to  the resonance frequency of the nanorods \cite{chen2014modulation,biagioni2012nanoantennas}. This resonance is  considered as a ``parametric resonance'' as it originates from the detuning of the resonance frequency due to varying structural parameters.
\begin{figure*}[ht]
	\includegraphics[width=0.8\linewidth]{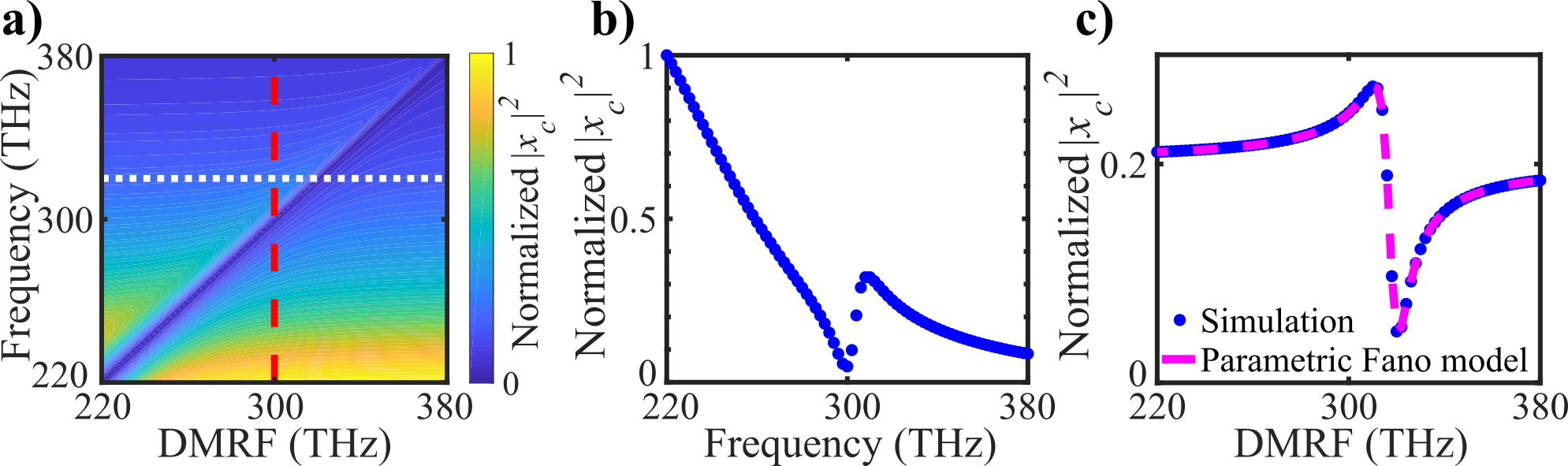}
	\caption{\label{fig2} Spectral and parametric Fano resonance in a CHO system. (a) Normalized simulated amplitude ($|x_c|^2$, color bar) of a system consisting of damped CHO mapped on the two-dimensional plane at the excitation frequency and the DMRF. (b) Simulated spectral Fano profile of the system at a fixed DMRF of 300 THz ( profile along the red dashed line in a). (c) Simulated (dotted blue line) and analytically calculated (dashed magenta line) parametric Fano profiles of the system at a fixed excitation frequency of 320 THz (profile along the white dotted line in a). The dashed magenta profile is obtained with Eq.\ref{eq6}. The parameters used are $\omega_{oc}$=220 THz, $\gamma_c$=144 THz, $\gamma_d$=5 THz, $g$=12000 THz$^2$, $A_1=1$ and $A_2=0$.}
\end{figure*}

The linear frequency response of the nanorod near the resonance frequency is well described by a simple harmonic oscillator as \cite{zuloaga2011energy},
\begin{equation}
	\ddot{{x}}+{\gamma}\dot{{x}}+{\omega}_{o}^{2}{x}={A}{E}_{o}{e}^{-{i\omega t}}
\end{equation}
Here, $E_oe^{-i\omega t}$ is the electric field of the time-harmonic excitation at frequency $\omega$, $x$ is the displacement for electrons from the mean position due to electric field, $\gamma$ accounts the damping of plasmon oscillation with all possible mechanism, $\omega_o$ is the resonance frequency that depends on the dimension of the structure and the material properties, and $A$ is a constant related to the total number of free charges and their effective mass. It is worth mentioning that the above equation can model the plasmon response of a particle in the quasistatic approximation and Drude’s regime of the material permittivity. The steady-state solution of the above differential equation is a Lorentzian resonance with amplitude ($\alpha$) and additional phase change ($\phi$), respectively, as,
\begin{equation}
	{\alpha}=\left|{x}\right|=\left|\frac{{A}{E}_{o}}{\left({\omega}_{o}^{2}-{\omega}^{2}-{i\gamma\omega}\right)}\right|
	\label{eq2}
\end{equation}
and
\begin{equation}
	{\phi}={{tan}}^{-{1}}{\frac{{\gamma\omega}}{{\omega}_{o}^{2}-{\omega}^{2}}}
	\label{eq3} 	              
\end{equation}
Eqs. \ref{eq2} and \ref{eq3} show that the amplitude and the phase of a given oscillator depend on the frequency of the driving force $\omega$. In a typical case of spectroscopic analysis, a broadband source is used to excite a nanostructure with a resonance frequency at $\omega_o$ around which the amplitude and phase profiles reveal the optical response of the oscillator in the frequency domain, i.e., the resonance spectrum. However, Eqs 2-3 also imply the possibility to describe the monochromatic optical response (at fixed $\omega$) of a series of resonators whose resonance frequency ($\omega_o$) depend on the variation of the structural dimensions such as length, height, and diameter \cite{chen2014modulation}. For example, the resonance frequency of a plasmonic nanorod depends on the nanorod length/aspect ratio. Therefore, the single-frequency amplitude and phase profiles in the structural parameter space also reveal the characteristics of the optical response of the oscillator. It is important to note that the phase profile in the frequency domain and the parametric resonance profiles shows ``opposite behavior'', i.e. the electric field shows a phase gain (lag) with increasing $\omega$ ($\omega_o$). 
\section{Parametric Fano resonance based on coupled driven damped harmonic oscillator}
In the following, we apply the concept of ``parametric resonance'' to describe the Fano resonance of a CHO system consisting of continuum (C) and a discrete (D) modes coupled through a coupling constant ($g$). In general, an electromagnetic field can be used to drive both the modes present in the Fano resonant system which is usually the case when both the modes are contained in the same structure like the oligomers or ring-disk cavity \cite{luk2010fano}. Assuming a linear frequency response of the two modes, the equation of motion of the system is \cite{gallinet2013plasmonic,oliver2013mechanisms},
\begin{equation}
	\begin{aligned}
		{\ddot{{x}}}_{c}+{\gamma}_{c}{\dot{{x}}}_{c}+{\omega}_{{oc}}^{2}{x}_{c}+{g}{x}_{d}={A}_{1}{E}_{o}{e}^{-{i\omega t}}  \\       
		{\ddot{x}}_d+\gamma_d{\dot{x}}_d+\omega_{od}^2x_d+gx_c={A}_{2}{E}_{o}{e}^{-{i\omega t}}
	\end{aligned}
\end{equation}	
Here, $\gamma_c$ ($\gamma_d$) and $\omega_{oc}$ ($\omega_{od}$) are the damping (with all possible mechanism) and the resonance frequency of the continuum (discrete) mode. The steady-state intensity response for the continuum mode can be written as:
\begin{equation}
	\left|{x}_{c}\right|^{2}=\left|\frac{A_1{E}_{o}\left({\omega}_{{od}}^{2}-{\omega}^{2}-{i}{\gamma}_{d}{\omega}\right)-gA_2{E}_{o}}{\left({\omega}_{{oc}}^{2}-{\omega}^{2}-{i}{\gamma}_{c}{\omega}\right)\left({\omega}_{{od}}^{2}-{\omega}^{2}-{i}{\gamma}_{d}{\omega}\right)-{g}^{2}}\right|^{2} 
	\label{eq5}
\end{equation}
$\left|x_c\right|^2$ gives the spectral Fano resonance of the CHO for a fixed value of all other structural parameters. Now, we consider a case where a driving force of fixed frequency applied on the continuum mode, and the discrete mode resonance frequency (DMRF) is tuned by, for example, varying the structural parameters of the coupled oscillators. For the cases where tuning the structural parameters does not significantly vary the damping of the two modes, the resonance frequency of the continuum and the coupling strength, $\omega_{od}$ is the only variable and Eq.\ref{eq5} becomes a function of $\omega_{od}$. Consequently, for a very dark discrete mode ($A_2\ll A_1$) \cite{limonov2017fano,joe2006classical,gallinet2013plasmonic}, $\left|x_c\right|^2$ can be approximated to give a parametric Fano resonance formula for a fixed excitation frequency,
\begin{equation}	
	\left|{x}_{c}({\omega}_{{od}})\right|^{2}\cong\left|\frac{{A}{E}_{o}}{\left({\omega}_{{oc}}^{2}-{\omega}^{2}-{i}{\gamma}_{c}{\omega}\right)}\right|^{2}\ \ \frac{\left({q}+{\eta}\right)^{2}+{b}}{\left({1}+{\eta}^{2}\right)}
	\label{eq6}
\end{equation}
Here, $\left|AE_o/\left(\omega_{oc}^2-\omega^2-i\gamma_c\omega\right)\right|$ is a constant for a given excitation frequency and the $\left(\left(q+\eta\right)^2+b\right)/\ \left(1+\eta^2\right)$  term is responsible for the parametric Fano resonance. $q=(\omega^2-\omega_{oc}^2)/(\gamma_c\omega\left(1+\Gamma_i/\Gamma_c\right))$ is the asymmetry parameter of the resonance. $\eta=(\omega^2-\omega_{od}^2-\omega_{od}\Delta)/(\Gamma_i+\Gamma_c)$ is the reduced energy scale with $\Delta=\Gamma_c(\omega^2\ -\omega_{oc}^2)/\omega^2\gamma_c$. $b=\Gamma_i^2/\left(\Gamma_i+\Gamma_c\right)^2$ is responsible for the Lorentzian background due to optical energy dissipation with $\Gamma_i=\gamma_d\omega$ and $\Gamma_c=g^2\gamma_c\omega/\left(\left(\omega^2\ -\omega_{oc}^2\right)^2+{(\gamma}_c\omega)^2\right)$. Eq.\ref{eq6} shows that a Fano CHO system driven by monochromatic excitation can exhibit a Fano profile in its optical response when the resonance frequency of the DMRF ($\omega_{od}$) is gradually tuned by varying a critical experimental or structural parameter. Varying the critical parameter tunes the DMRF, thus changing the amplitude and phase relationships between the two interfering modes. This is similar to scanning the excitation frequency in a typical case of spectral analysis in the frequency domain. 

In Fig. \ref{fig2}, we use MATLAB to evaluate the amplitude response of the coupled oscillators using Eq. \ref{eq5} for varying excitation frequency ($\omega$) and DMRF ($\omega_{od}$), and the other oscillator parameters mentioned in the caption. The amplitude response of the continuum mode of a CHO is mapped on a two-dimensional plane with respect to the excitation frequency ($\omega$) and the DMRF ($\omega_{od}$) related to a tuned experimental parameter. The sharp variation between the distribution stems from the Fano interference of the CHO. Fig.\ref{fig2}b shows the typical frequency-domain Fano spectrum of a CHO system with fixed experimental parameter, i.e. a fixed DMRF ($\omega_{od}$= 300 THz), excited by a broadband source. This corresponds to the intensity profile along the red dashed line in Fig.\ref{fig2}a. Similarly, we can fix the excitation frequency at 320 THz (white dotted line in Fig.\ref{fig2}a) and plot the monochromatic response intensity of the CHO system with respect to the tuned DMRF. In this way, we obtain a ``parametric'' Fano resonance profile (Fig.\ref{fig2}c), which is indeed in the parametric domain since the DMRF can be associated with a critical experimental parameter. Such parameter can be a critical non-spectral experimental variable, like temperature, polarization, or structural dimensions, for example, a chirped periodicity of a grating  \cite{chen2021spectrometer,see2017design,ouyang2020spatially}. Eq.\ref{eq6} is not only useful in describing the intensity profile of a parametric Fano resonance, it also provides information about the dependence of the asymmetry on the critical experimental or structural parameters, which is important for the design of Fano nanoresonators for monochromatic applications. It is worth noting that due to the phase behavior of the Lorentzian mode in the parametric space (Eq.\ref{eq6}), the asymmetricity of the Fano profiles in spectral (Fig.\ref{fig2}b) and parametric (Fig.\ref{fig2}c) domains are opposite, i.e., the $\left|x_c\right|^2$ maxima of the two Fano resonance profiles appear on different sides of intensity minima.
\begin{figure}[t]
	\centering
	\includegraphics[width=0.9\linewidth]{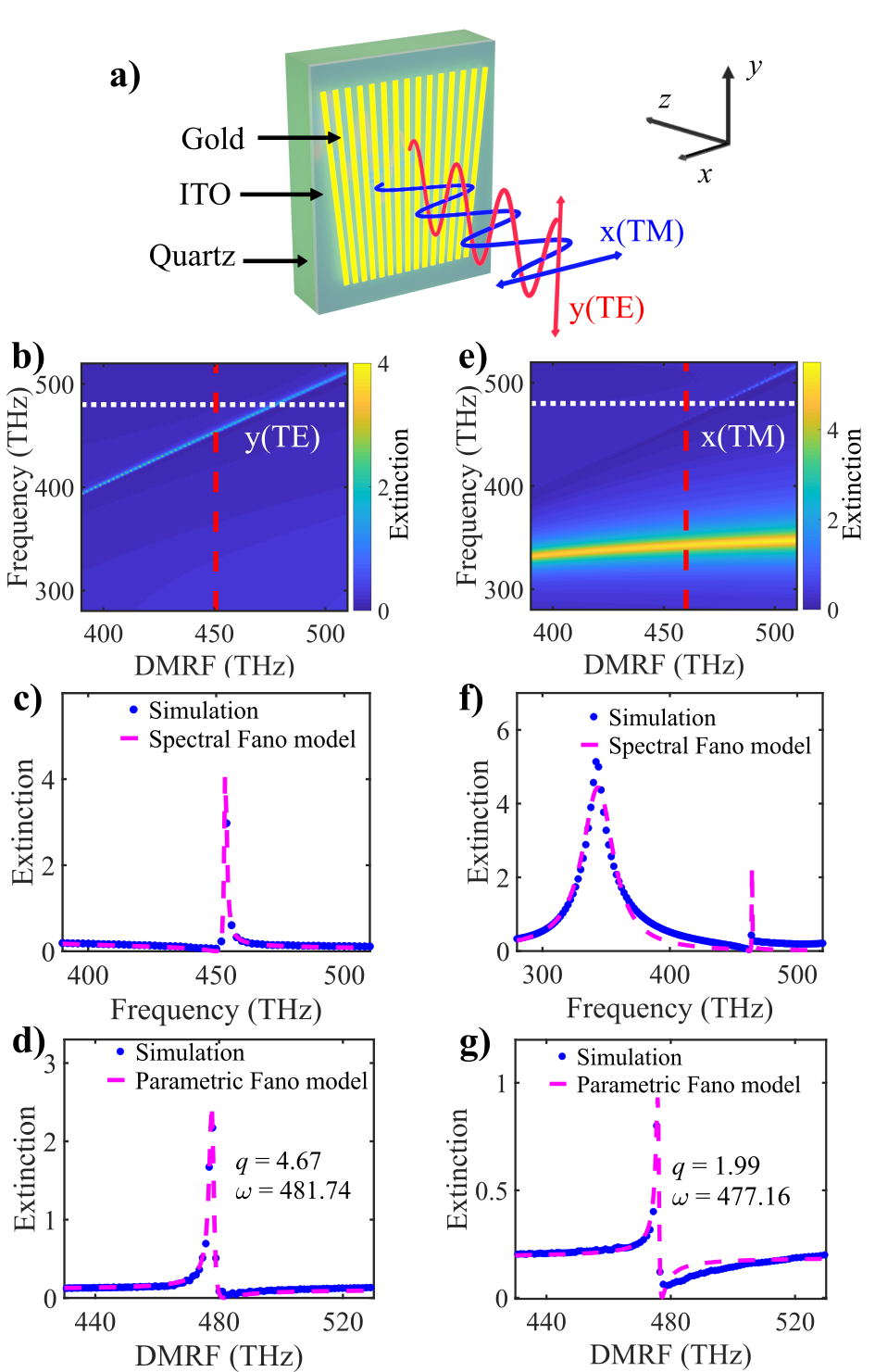}
	\caption{Parametric resonances on the extinction cross-section of the CWPC for input TE and TM polarization.  (a) Schematic of the CWPC that exhibits Fano resonance in the extinction cross-section.  (b) The TE-polarized extinction cross-section (color bar) of CWPC mapped on the two-dimensional plane of the excitation frequency and DMRF. (c) TE polarization extinction cross-section spectrum at a given DMRF (along the red dashed line in Fig. b), the dashed magenta curve shows a fit with Eq.\ref{eq5}. (d) DMRF dependence of the TE polarization extinction cross-section at a fixed excitation frequency (along the white dotted line in Fig. b), the dashed magenta curve indicates a fit with Eq.\ref{eq6}. (e) The two-dimensional distribution of extinction cross-section (color bar) of WPC for TM polarization. (f) TM polarization extinction cross-section spectrum at a given DMRF, (along the red dashed line in Fig. e), the dashed magenta curve shows a fit with Eq.\ref{eq5}. (g) The DMRF dependence of the extinction cross-section of TM polarization (along the white dotted line in Fig. b), the dashed magenta curve indicates a fit with Eq.\ref{eq6}.}
	\label{fig3}
\end{figure}

 Now, we demonstrate the parametric Fano model on a CWPC, which consists of a gold grating with a spatially chirped period ($d_{x}$) that varies along the y-axis. The grating is placed on  a 160-nm thick Indium Tin Oxide (ITO, $n_{ITO}$=1.8) coated quartz substrate ($n_{sub}=1.5$), as shown in Fig.\ref{fig3}a. The height and width of the plasmonic nanowires are set to be 15 nm and 150 nm, respectively. In this system, the spectral Fano resonance in the CWPC originates from the spectral interference of the discrete waveguide mode (bounded mode excited by the grating) in the ITO layer with the broad plasmon mode of the grating for incident light with transverse magnetic (TM) polarization \cite{miroshnichenko2010fano,luk2010fano,christ2004PRB}. Under transverse-electric (TE) illumination, the incident light with a broadband spectrum acts as the continuum mode, interferes with the discrete TE waveguide mode, resulting in Fano resonance \cite{miroshnichenko2010fano,luk2010fano,christ2004PRB}.
\section{The parametric Fano Resonance in CWPC}
In order to obtain the optical response of CWPC, we simulate (FDTD Solutions, Lumerical, see Appendix) extinction spectrum (defined as -log (transmission)) of individual WPCs of a fixed period and merge the spectrum for different periods to obtain the complete optical response of the CWPC. The grating and the waveguide area is meshed with square meshes of size 0.25 nm$^2$. It is noteworthy that the used simulation method for CWPC assumes that the length of CWPC in the y-direction is much larger than its width in the x-direction. Fig.\ref{fig3}b shows the simulated (FDTD Solutions, Lumerical) extinction cross-section, defined as -log (transmission), of the CWPC on the two-dimensional plane for the excitation frequency and resonance frequency of the discrete grating mode, which can be linked to the period varying  along the y-coordinate. The DMRF is obtained from the grating period using the dispersion relation of the waveguide mode \cite{kogelnik1974scaling} (see Appendix). Fig.\ref{fig3}c shows the TE extinction spectrum with a distinct asymmetric spectral Fano line shape in the excitation frequency domain, with a fixed DMRF at 450 THz, from a grating period of 430 nm. The simulated spectrum is fitted by the spectral Fano model shown in Eq.\ref{eq5}. The fitting parameters are given in Appendix table S1. Turning to the parametric domain, Fig.\ref{fig3}d shows the monochromatic extinction cross-section (excitation frequency = 480 THz) as a function of the spatially chirped DMRF. As a function of the DMRF, a clear parametric Fano-like profile is obtained, which depends on the chirped periodicity of the grating. The simulated profile of the parametric domain can be well fitted by the developed parametric Fano model (Eq.\ref{eq6}.), validating our parametric Fano model.  The fitting parameters are given in Appendix table S2. 

The response of the CWPC to a TM excitation is also studied. As shown in Fig.\ref{fig3}e, a similar Fano resonance in the extinction spectrum is observed, despite its different origin. The strong Lorentzian resonance peak of the plasmon continuum mode ($\approx$340 THz) interferes with the discrete TM waveguide mode, giving an asymmetric Fano resonance line shape in the spectral domain. Fig.\ref{fig3}f shows the frequency-domain extinction spectrum of the grating with a fixed period of 430 nm, corresponding to a resonance frequency of 460 THz for the discrete mode. The spectral Fano model fit (Eq.\ref{eq5}) is used to check the validity of the model on TM polarized extinction spectrum. The fitting parameters are given in appendix table S1. Fig.\ref{fig3}g shows the parametric Fano resonance profile for a TM polarized excitation with the excitation frequency fixed at 480 THz. The resonance profile can be well fitted using the parametric Fano resonance model in Eq.\ref{eq6}. The fitting parameters are given in appendix table S2. The fit gives a quantitative estimate of the asymmetry of resonance ($q$) and excitation frequency ($\omega$).  
\section{Conclusion}
The concept of parametric Fano resonance maps the spectral interference phenomenon of the Fano resonance to different domains of experimentally accessible parameters. It is to be noted that such parametric resonances are not exclusive for the CWPC presented here, but could also be implemented by varying relevant material or other parameters of the system, such as refractive index, orientation, and morphology of the structure. Some of these parameters might even be controlled externally by tuning the system temperature \cite{zhang2018thermally,zheng2017compact} or the polarization of light \cite{ray2017polarization,valentim2013asymmetry}. In addition to the fundamental interest in understanding Fano resonances observed in the parametric domains, the developed model may find applications in monochromatic nanophotonic applications, where Fano-like nanoresonators are to be optimized for a specific operational frequency, for example, plasmonic index sensors and nanostructures for Raman or other nonlinear signal generations.

    \section{Acknowledgments}
    Financial supports from the Deutsche Forschungsgemeinschaft (CRC 1375 NOA, HU2626/3-1, HU2626/5-1) and the Ministry of Science and Technology of Taiwan (MOST-103-2113-M-007-004-MY3) are gratefully acknowledged.
\appendix    
    \begin{center}
        \vspace{1cm}
        \textbf{\large Appendix} 
    \end{center}
    \section{The dispersion relation of transverse electric (TE) and transverse magnetic (TM) modes}
    The dispersion relations of the waveguide modes in the 160 nm thick ITO layer was calculated by solving the transcend equations for a system without the gold grating.  The solution gives an effective refractive index of the waveguide mode for various frequency of the incident electromagnetic waves. The effective refractive index was used to find the momentum matching condition of the waveguide mode and the momentum provided by the grating. Figure S1 shows the the discrete mode resonance frequency (DMRF) as a function of grating periodicity obtained from the momentum matching condition for TE and TM modes. The curve truncates once the cut off frequency of the waveguide mode is reached at the lower frequencies.
    \begin{figure}[b]
	\centering
	\includegraphics[width=0.7\linewidth]{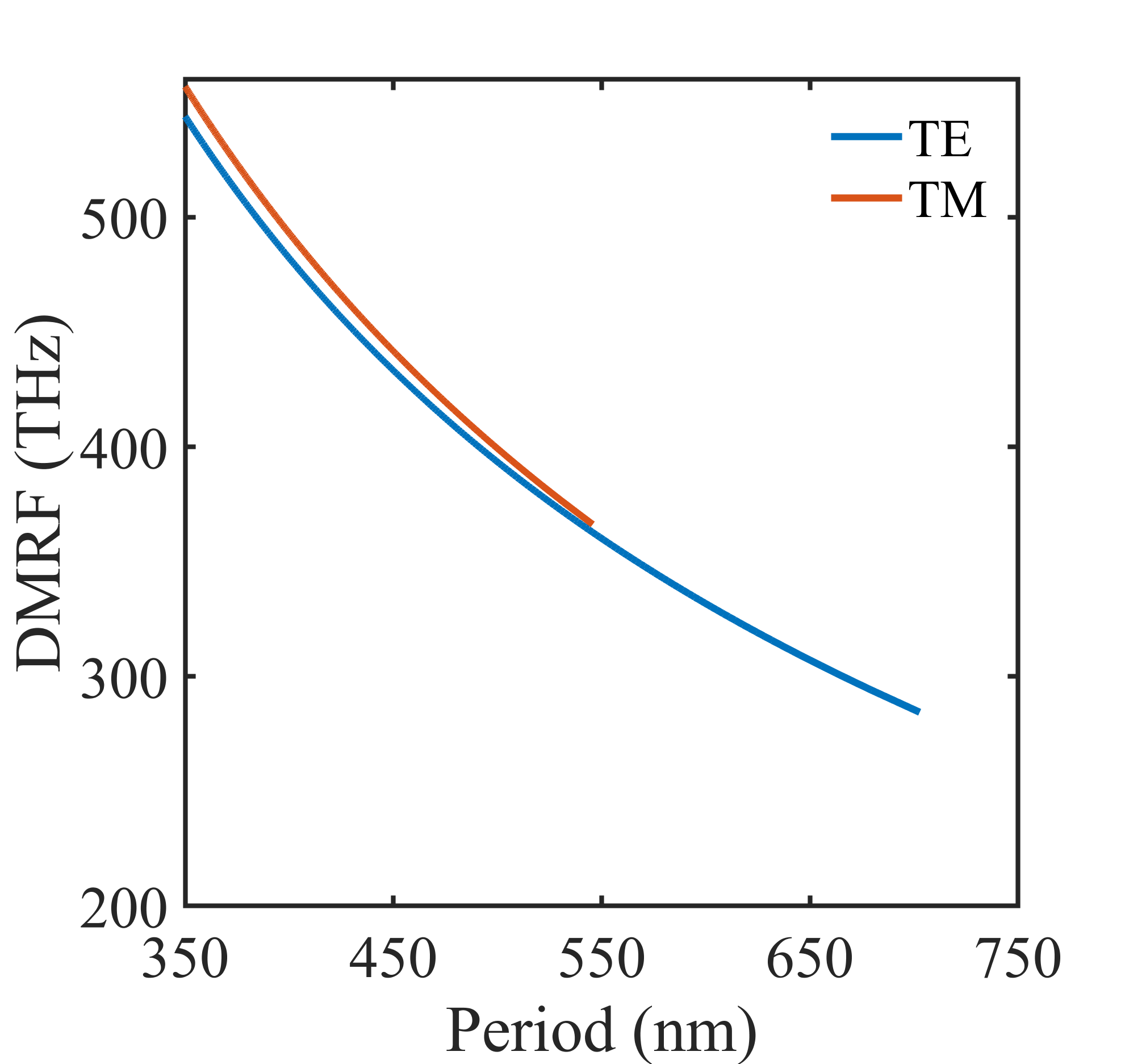}
	\caption{The dispersion relation showing the dependence of resonance frequency of TE and TM waveguide mode in the ITO layer on the period of grating used to excite the mode.}
	\label{figS1}
\end{figure}
\section{FDTD simulation method}
\subsection{Scattering form nanorod} A TFSF source in the Lumerical FDTD was used to simulate the scattering cross section of the nano rod. The polarization of light was kept parallel to the long axis of nanorod. The scattering cross-section was recorded using the cross-section analysis group provided by Lumerical. In order to record the phase spectrum a point monitor was placed at the centre of nanorod. The boundary condition was set to PML and the nanorods were meshed with 1 $nm^3$ mesh size. The Simulation region was kept approximately of size of $5.8 \mu m^3$ with PML layer kept at the distance of at least 900 nm from the surface of nanorod. The simulation was terminated only after reaching the auto shutoff value less than $10^{-5}$. The simulation results of different nanorod lengths were combined together to obtain the scattering cross section of gold nanorod on a two-dimensional plane of the excitation frequency (vertical axis) and the nanorod length (horizontal axis), shown in figure 1a.

\subsection{Scattering form Grating}The scattering response from chirped waveguided plasmonic crystal (CWPC) was obtained by simulating the individual gratings of different periodicities and the results of the simulation of individual grating period were combined together to obtain the optical response of CWPC. This method of simulation assumes that for the x-(y-) polarized incidence, the contribution from TE (TM) polarization can be neglected. The assumption is valid if the length (along y) of the chirped grating is very large compared to its width (along x, direction of grating periodicity) or in other words when we can neglect the tilt of grating in y direction for a given period.

To simulate the response of the grating with fix period, a 2D FDTD domain was set up to simulate scattering from an infinite plasmonic grating illuminated by a plane wave source from the air side. A field monitor was placed on the other side of grating in the substrate medium to record the transmission. The grating and the waveguide area was meshed with $0.25 nm^2$ mesh size. A periodic boundary condition was placed in the direction of grating period (x-axis), while a PML boundary condition was applied on the y-direction at a distance of approximately 950 nm from the grating. The simulation was terminated only after reaching the auto shutoff value less than $10^{-5}$.

\begin{figure*}[b]
	\centering
	\includegraphics[width=0.9\linewidth]{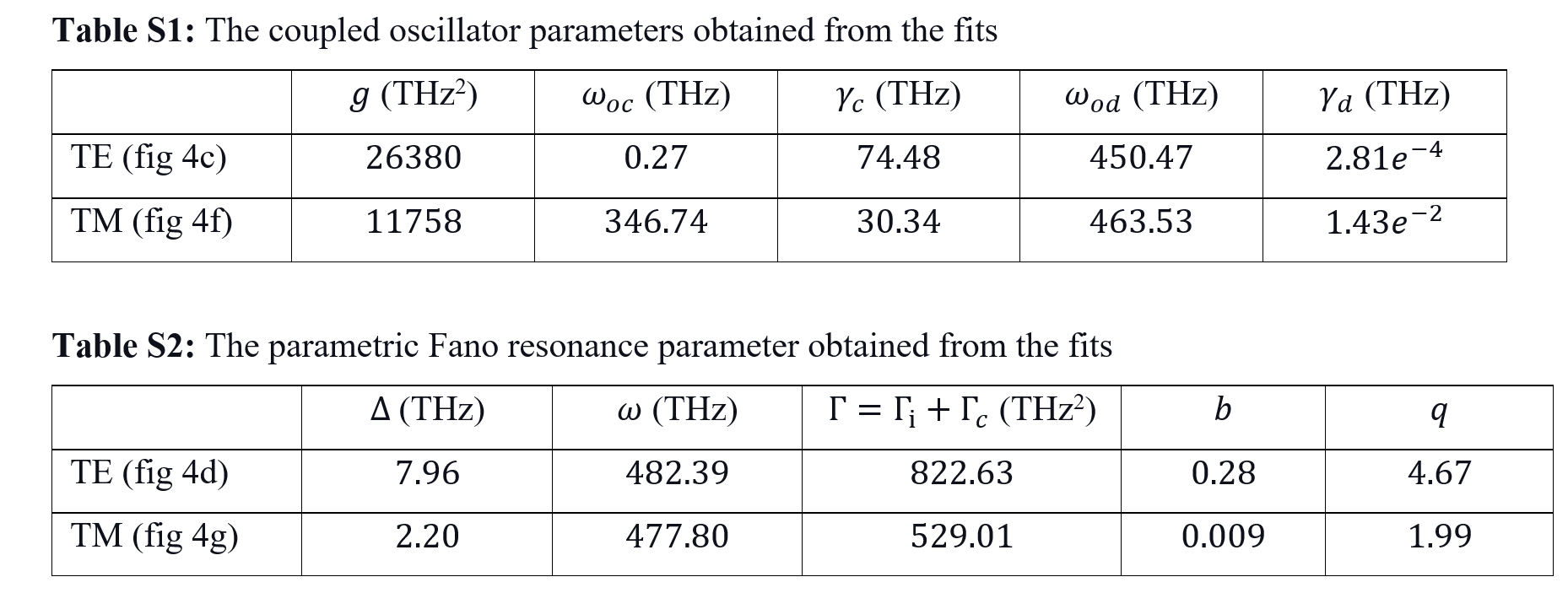}
	\label{figS2}
\end{figure*}

\bibliographystyle{apsrev4-1}
\bibliography{ref.bib}   
\end{document}